\documentclass[reprint,superscriptaddress]{revtex4-1}
\usepackage{blindtext}
\usepackage{placeins}
\usepackage{nicefrac}
\usepackage{siunitx}
\usepackage{setspace}
\usepackage{textcomp}
\usepackage{graphicx}
\usepackage[normalem]{ulem}

\begin{document}
	
	\newcommand{\inst}{Institut f\"ur Angewandte Physik, Universit\"at Bonn, Wegelerstr. 8, 53115 Bonn, Germany}
	\newcommand{\insti}{Present address: Complex Photonic Systems (COPS), MESA+ Institute for Nanotechnology, University of Twente, PO Box 217, 7500 AE Enschede, Netherlands}
	\newcommand{\instii}{Present address: Cavendish Laboratory, University of Cambridge, 19 JJ Thomson Avenue, Cambridge CB3 0HE, United Kingdom}

	\title{Realizing arbitrary trapping potentials for light via direct laser writing of mirror surface profiles}
	
	\author{Christian Kurtscheid\textsuperscript{1}, David Dung\textsuperscript{1}, Andreas Redmann\textsuperscript{1}, Erik Busley\textsuperscript{1}, Jan Klaers\textsuperscript{1,*}, Frank Vewinger\textsuperscript{1}, Julian Schmitt\textsuperscript{1,**}, and Martin Weitz}
\affiliation{\inst\\
	\textsuperscript{*}\insti \\
	\textsuperscript{**}\instii}
	
	\begin{abstract}
	The versatility of quantum gas experiments greatly benefits from the ability to apply variable potentials. Here we describe a method which allows the preparation of potential structures for microcavity photons via spatially selective deformation of optical resonator geometries with a heat induced mirror surface microstructuring technique. We investigate the thermalization of a two-dimensional photon gas in a dye-filled microcavity composed of the custom surface-structured mirrors at wavelength-scale separation. Specifically, we describe measurements of the spatial redistribution of thermal photons in a coupled double-ridge structure, where photons form a Bose-Einstein condensate in a spatially split ground state, as a function of different pumping geometries. 
	\end{abstract}
	
	\maketitle

	\section{Introduction}
	\vspace{-0.2cm}
	The strength of manipulating and controlling quantum gases to observe a variety of quantum physics phenomena relies critically on the ability to confine the ensembles in variable potentials. Periodic lattice potentials, as a prime example, have been realized for ultracold atom systems, exciton-polaritons, and more recently also for Bose-Einstein condensates of photons\cite{atomLattice1, Abbarchi,berloff,Ohadi,David}. However, non-periodic potentials are harder to prepare, especially in cold atom systems.
	
	In this letter, we discuss a novel technique to apply arbitrary potentials for photon quantum gases in optical microcavities. Bose Einstein condensation of photons has been realized for the first time in a dye-filled optical microcavity \cite{JanCondensed,Nyman0,Dries}, where (in contrast to blackbody radiation) the chemical potential of the photons does not vanish,  and an independent control of photon number and temperature is achieved. In this geometry, the small, wavelength-sized mirror spacing introduces a low-frequency cutoff and makes the dispersion quadratic, i.e. matter-like, and the system two-dimensional. Thermal equilibrium is achieved by repeated absorption re-emission processes of the photons on the dye molecules, realizing a thermal contact to the room-temperature dye solution. For this, high mirror reflectivities are crucial to provide long cavity photon storage times, which exceed the required equilibration time of the photon gas.
	
	The microcavity geometry, i.e. transverse variations in the optical path between the mirrors, can provide potentials for the two-dimensional photon gas. At transverse positions with larger spacing between reflecting mirror surfaces, for example, longer wavelength (lower frequency) photons match the mirrors’ boundary condition, giving rise to a spatial dependence of the photon potential energy \cite{Kurtscheid,JanJosephson}. Modifying the optical path via a local refractive index gradient has been successful in the realization of tight harmonic traps and the observation of coupled photon condensates in a double-well \cite{David}. Physically imprinting concave curvatures to the mirror substrate via focussed ion beam milling prior to dielectric coating has led to the observation of condensates formed by a few photons in deep harmonic confinements \cite{trichet2015topographic,Nyman1}. More recently, we have presented a novel technique with which Bragg mirrors can be shaped transversally by local heating of the dielectric layers. Implemented in the microcavity setup, it enables rapid prototyping for a wide range of potential landscapes and has already led to the realization of a photon condensate in a coherently split superposition state \cite{Kurtscheid}. Here, we present a detailed description of the structuring technique and discuss thermalization properties of the coherently split photon gas when subject to nonequilibrium state preparation. 	
	\begin{figure}[t]
		\centering
		\includegraphics[width=0.49\textwidth]{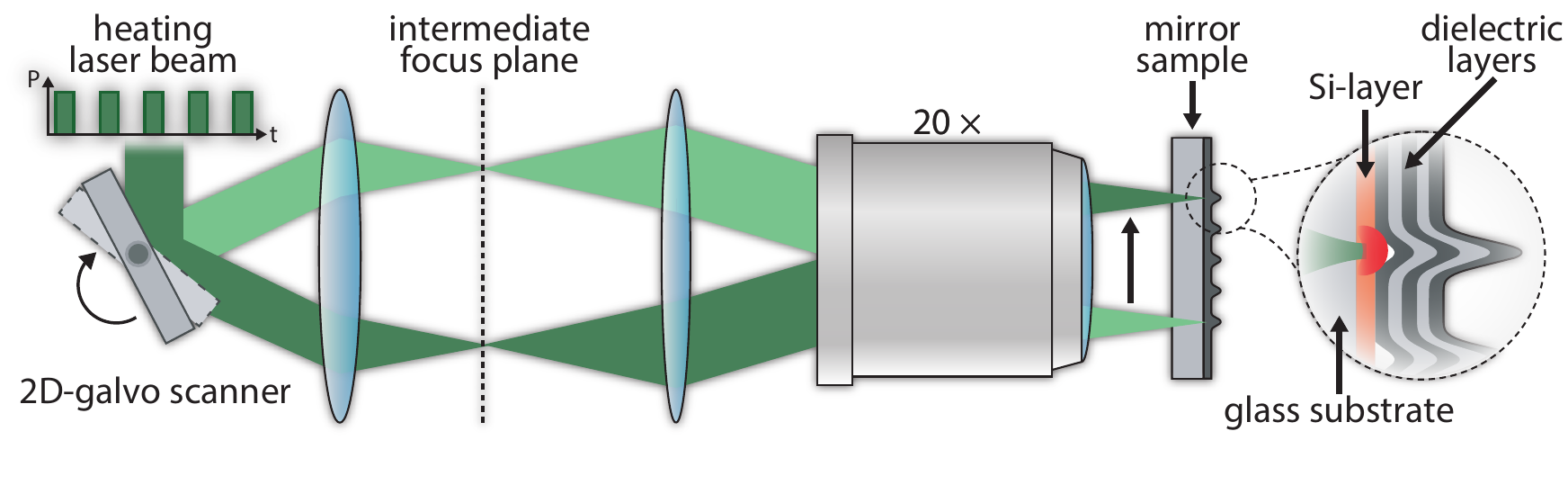}
		\caption{Schematic of the mirror microstructuring setup. A power-modulated heating laser beam is deflected in two dimensions by a galvo scanner. A lens system equipped with a high NA objective maps the deflection angle to the position of the focal spot on the sample. In a thin silicon film between glass substrate and dielectric layers the laser beam is partially absorbed, resulting in local heating and permanent deformation of the mirror surface (height not to scale).  }
		\label{fig.1}
	\vspace{-.2cm}
	\end{figure}
	\vspace{-.2cm}
	\section{Controlled heat-induced mirror structuring} 
	
	For our mirror structuring technique we use ultra-high-reflectivity Bragg-mirrors, which are coated with an additional \SI{30}{nm} film of silicon between the glass substrate and the dielectric layers, as writing samples (coating manufacturer Laseroptik, center wavelength \SI{550}{nm}). As depicted in the schematic setup in Fig. 1, a laser beam at \SI{532}{nm} wavelength is temporally power-modulated by an acousto-optic modulator and transversally steered using a two-dimensional galvo scanner system. The beam deflection angle is translated into a localized spot in the intermediate focal plane by a lens and subsequently projected onto the mirror sample from the reverse side via a high NA microscope objective. By controlling both laser intensity and transverse position on the mirror, a spatio-temporally variable heat pattern can be induced in the dielectric layers as the silicon absorbs $\sim$35\,\% of the incoming laser light. For sufficiently large optical powers, the heating is found to result in a permanent deformation of the mirror surface.
	
	Figure 2a exemplarily shows the surface height profile after heating with a 5×5-point-pattern determined in an interferometric measurement where the optical setup including a Mirau objective (NA=0.4) and a \SI{530}{nm} center wavelength LED as light source enables a lateral and longitudinal resolution of \SI{0.74}{\micro m} and \SI{0.1}{\angstrom} respectively (see ref. \cite{mirau} for a similar setup). Although the width of the heating laser beam is only \SI{1.8}{\micro m} (FWHM) at the surface of the mirror, the observed Gaussian deformation exhibits a power-independent intrinsic width of \SI{3.6}{\micro m}. We only observe measurable surface changes above a threshold power of around \SI{25}{mW}, see Fig. 2b. Further increasing the power leads to a steadily growing Gaussian deformation of typically up to \SI{50}{nm} height at a power of \SI{70}{mW}. Beyond this value, the outcome is observed to become unpredictable (see error bars), as the resulting height varies steeply from \SI{100}{nm} to \SI{230}{nm} until at even higher powers a saturation is observed. Measurements here reveal a significant increase of the mirror transmission at a wavelength of \SI{532}{nm} from few ppm to a single-digit percentage, indicating the destruction of the Bragg layers beyond a resulting height of more than \SI{50}{nm}. The observed deformations might be caused by the successive delamination of the staggered dielectric layers from the underlying glass substrate, a phenomenon known to occur in stacks of thin layers where induced thermal stress can overcome the cohesive forces\cite{del1,del2}; yet without detailed measurements of the layer structure, which we plan to conduct with electron microscope scans, we can not confirm this hypothesis.
	
	To manufacture complex structures beyond single-Gaussian shapes, we use an iterative procedure where through multiple laser scans of the heating beam over the sample backside a desired height profile is imprinted. In the course of the writing process, the deviation between the current and the target profile is monitored and the spatial laser power profile is adapted in proportion to the measured local height difference. Figure 3 exemplarily shows the prototyping evolution of a mirror surface, where the target structure was set to feature a radius of curvature of \SI{15}{cm} with two additional minima aligned with the y-axis covering a total area of \SI{130}{\micro  m}$\times$\SI{130}{\micro m}. Starting from a flat mirror surface, at first a spatial calibration of the mirror dimensions is performed via the deformation of the profile edges. The laser beam power is then adjusted in several iterative cycles in order to gradually achieve the desired height profile. Figures 3b and c show the emergent surface structure and vertical cuts, respectively, after 7, 14 and 56 iterative cycles. As shown in Fig. 3d, the mean deviation from the target mask is significantly reduced down to \SI{1}{\angstrom}. This procedure, thus, is compatible with the requirements of super-polished optics. In order to quantify the impact of the delamination structuring procedure on the mirror quality we determined the reflectivity of imprinted flat-top surfaces with roughnesses near \SI{1}{\angstrom} and varying heights on an area of \SI{500}{\micro  m}×\SI{500}{\micro m} in cavity-ring-down measurements (cf. \cite{CRD}). We found that up to deformation heights of around \SI{18}{nm} the mirror reflectivity is unaffected within the experimental uncertainty, see the data shown in Fig. \ref{fig.3}e. A significant reduction of the measured mirror reflectivity is observed only at larger deformation heights, where as described above also transmission losses increase. As will be discussed in the following,  the reflectivity-conserving surface height profile modulations below \SI{18}{nm} are sufficient to form effective potential landscapes for photons in a dye-filled microcavity setup.  
		\begin{figure}[t]
	\centering
	\includegraphics[width=0.49\textwidth]{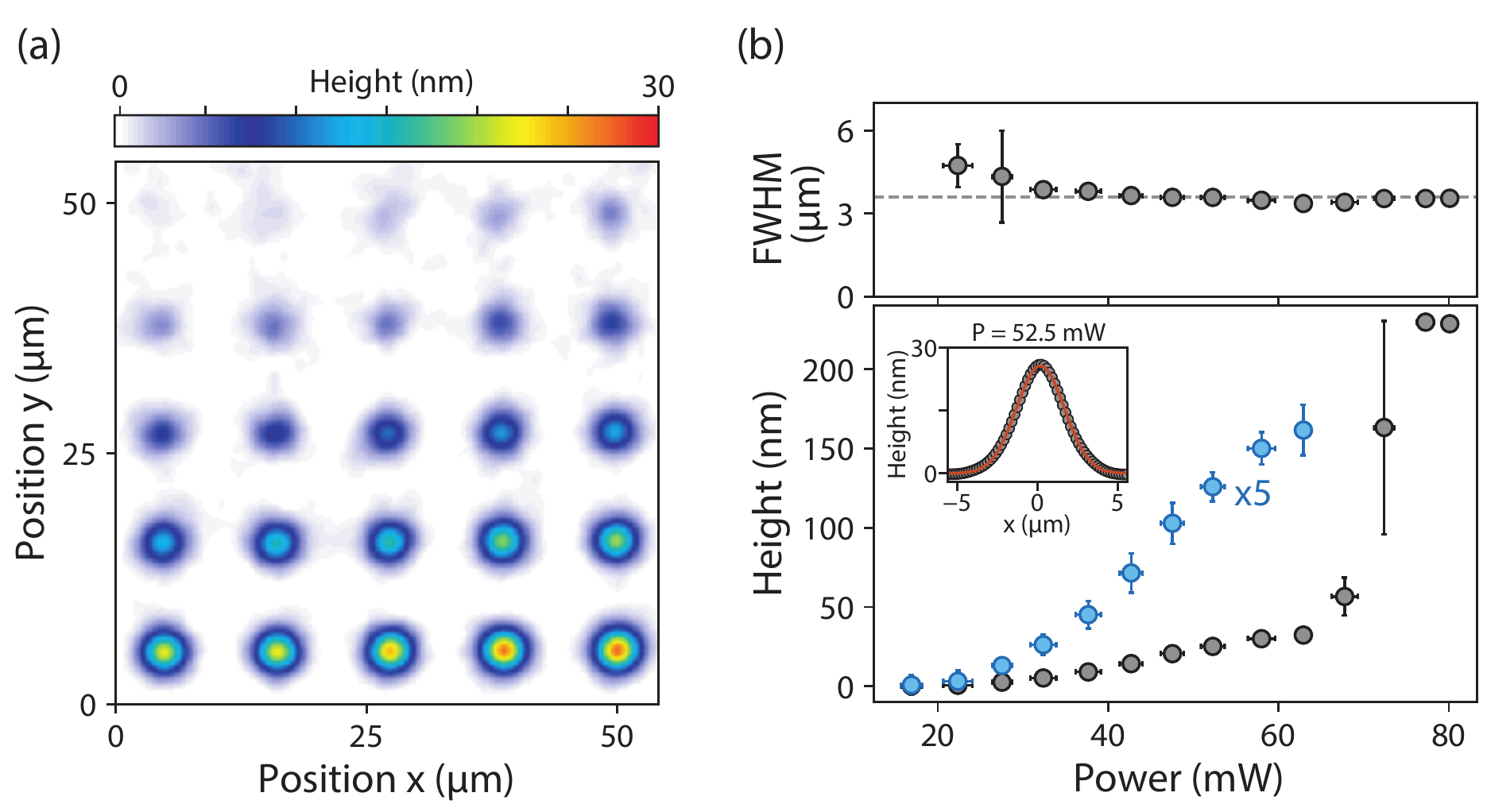}
	\caption{ (a) Example of a structured Bragg-mirror surface after point-like heating with laser beam powers from 25\,mW (top left) up to 54\,mW (bottom right). (b) Full width at half maximum (FWHM, top) and maximum height (bottom) of the induced deformation as a function of the heating laser power, both determined by Gaussian fits as shown in the inset. }
	\label{fig.2}
\end{figure}

	\section{Experimental applications}
	Bose-Einstein condensation of photons in a harmonic trap has been realized in a dye-filled bi-spherical microcavity at room temperature \cite{JanCondensed,Nyman0}. The short spacing of a few wavelengths between the two curved mirrors renders the photon gas effectively two-dimensional as the cavity free spectral range is in the order of the spectral width of the dye fluorescence and only one longitudinal mode can be energetically populated. More generally, assuming that the local transverse mirror distance $D(x,y)=D_0-h(x,y)$ shows small deviations $h(x,y)$ from the maximum mirror spacing $D_0$, the photon dispersion relation $E =\hbar c \sqrt{k_x^2+k_y^2+k_z^2 }$  , where $k_{x,y}$ are the transversal and $k_z$ the longitudinal component of the wave vector, can in the paraxial limit ($k_{x,y} \ll k_z$) be approximated by
	\begin{equation}
	E \simeq m_\mathrm{ph} \left(\frac{c}{n}\right)^2 + \frac{\hbar^2}{2m_\mathrm{ph}} (k_x^2 + k_y^2) + V(x,y).
	\end{equation}
	Here, the (effective) photon mass $m_\mathrm{ph}=nq\pi\hbar⁄(cD_0 )=2\pi \hbar n^2 /(c \lambda_\mathrm{c})$ is determined by the cutoff wavelength $\lambda_\mathrm{c}$ with $c$, $q$, and $n$ being the vacuum speed of light, the longitudinal wave mode number and the refractive index of the dye solution, respectively. The potential $V(x,y)$ scales linearly with $h(x,y)$ according to \cite{Kurtscheid,JanJosephson}
	\begin{equation}
	V(x, y)=m_\mathrm{p h} \frac{c^{2}}{n^{2} D_{0}} h(x, y).
	\label{eqV}
	\end{equation}
	Equation (\ref{eqV}) has been employed in earlier work, where curved mirrors provided a radially symmetric harmonic trap $V(x,y)\propto(x^2+y^2 )$, in which photon condensation to the Gaussian oscillator ground state has been observed and studied in detail \cite{JanCondensed,Julian1,Dries,Nyman0}. The thermalization of the photon gas was induced by frequent absorption and re-emission processes on the dye molecules being at thermal equilibrium. \\
	\begin{figure}[t]
		\centering
		\includegraphics[width=0.49\textwidth]{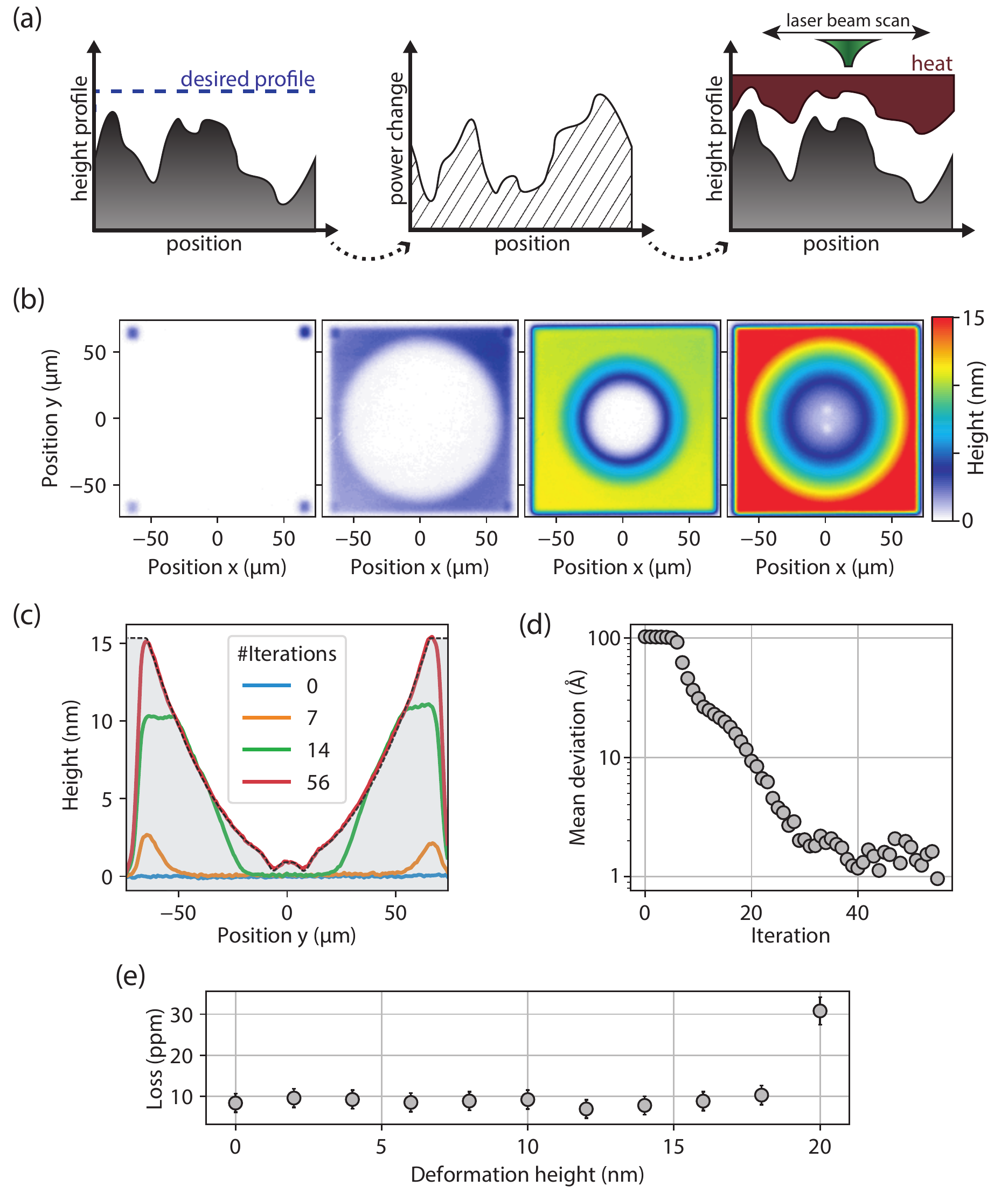}
		\caption{ Adaptive shaping of mirror surfaces. (a) Sketch of a single structuring cycle. From left to right: first, the height difference between the current and desired mirror profile is determined in a Mirau-interferometric surface measurement. In a second and third step, a heating beam corrective power pattern is calculated and applied to the existing structure. (b) Snapshots of the mirror surface prototyping, showing the spatial calibration step and iterations 7,14, and 56 (from left to right), realizing a surface mask with a radius of curvature near 15 cm with two additional indents along the y-axis. (c) Respective profile cuts along the vertical axis and target profile (dashed line). (d) Mean deviation between actual and target surface versus number of iterations. (e) Total mirror loss as obtained from a cavity-ring-down measurement as a function of the deformation height.}
		\label{fig.3}
	\end{figure}
	Under the same experimental conditions, the mirror structuring technique described above now enables the design of arbitrarily shaped potential landscapes for trapped photon gases, see also the schematics shown in Fig. \ref{fig.4}a. For this, let us consider a microcavity consisting of the spatially structured mirror of Fig. 3b and a plane cavity mirror, which induces a double well potential superimposed by a shallow harmonic trap; for details see  \cite{Kurtscheid}. Coherent tunneling of photons between the two minima results in the hybridization of the localized wavefunctions $\psi_1$ and $\psi_2$ in the (uncoupled) wells to symmetric and anti-symmetric superposition states 
	\begin{equation}
	\psi_\mathrm{s} = \frac{1}{\sqrt{2}} \left(\psi_1 + \psi_2\right),\ \ \  
	\psi_\mathrm{a} = \frac{1}{\sqrt{2}} \left(\psi_1 - \psi_2\right)
	\end{equation}
	respectively.  Here $\psi_\mathrm{s}$, the symmetric linear combination, has lower energy and forms the system ground state. The additionally imprinted harmonic trapping potential yields the required scaling of the density of states with energy to enable Bose-Einstein condensation in two dimensions. Experimentally, thermalization of photons within a similar potential has allowed to successfully produce macroscopic occupation of photons into the coherently-split lower-energy state, beyond the critical photon number of $N_\mathrm{c}$=8,000 \cite{Kurtscheid}. Moreover, the coherence of this thermodynamic beamsplitter was experimentally verified by overlapping the emission from the individual sites in an interferometric setup. In more recent work, a dye microcavity setup has also produced split states of light with variable relative phase, realizing a controllable Josephson junction \cite{JanJosephson}. For this study, both the here described permanent microcavity structuring method as well as the thermo-optic refractive index structuring technique from ref. \cite{David} were combined. 
	\section{Spatial condensate response to pump beam displacements}
				\begin{figure}[t]
		\centering
		\includegraphics[width=0.49\textwidth]{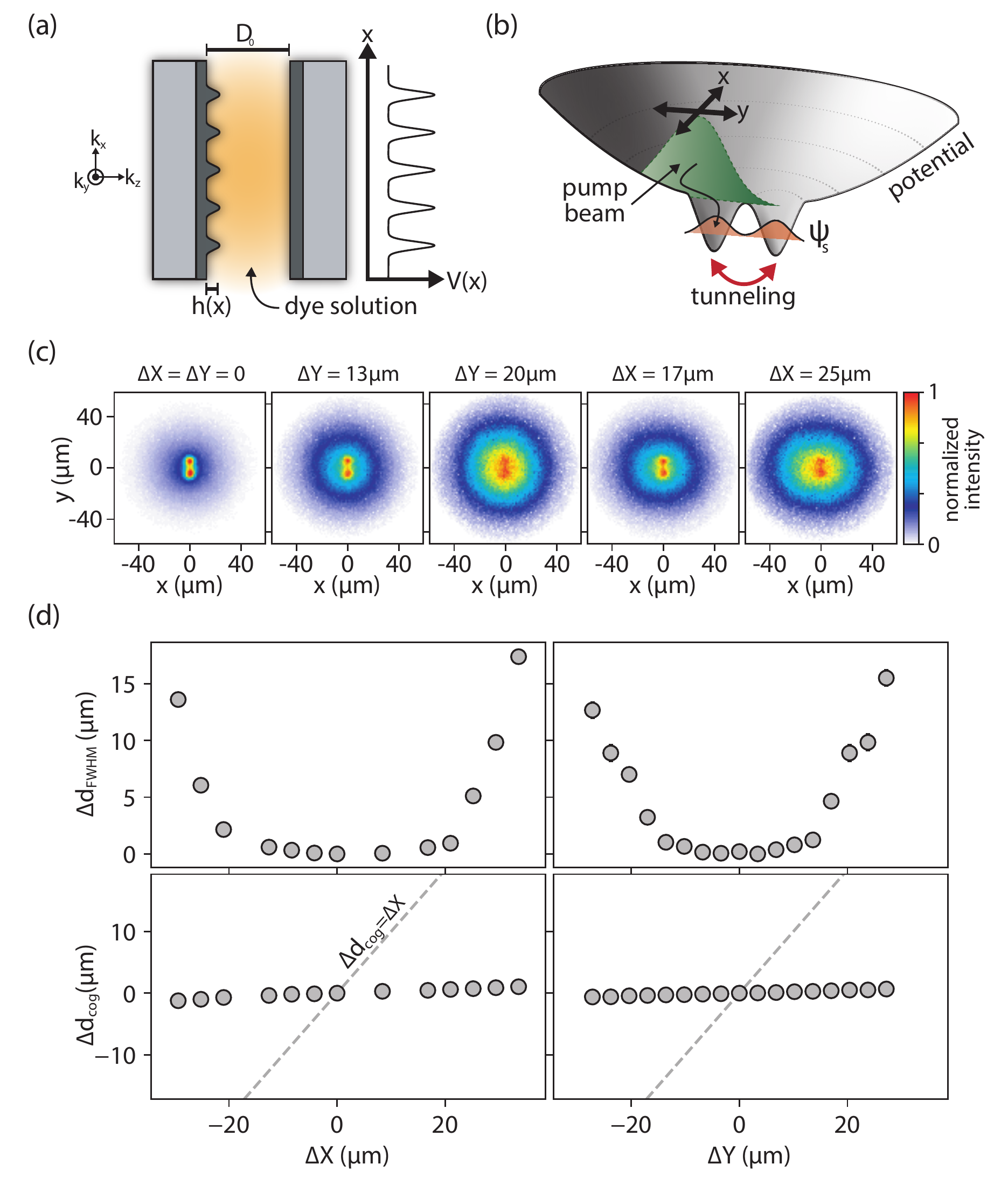}
		\caption{(a) In the paraxial limit, the local mirror surface height $h(x)$ linearly translates into an imprinted potential $V(x)$ for the photon gas in the dye-filled microcavity. (b) The pump beam is spatially steered over the double-well potential parallel and perpendicular, respectively, to the double-well axis. The thermalization process causes spatial redistribution of photons into the trap center. (c) Fluorescence images for varying pump beam positions. (d) Spatial width increase  $\Delta d_\mathrm{FWHM}$  for the thermal cloud (top) and position change of the center of gravity of the total emission $\Delta d_\mathrm{cog}$ (bottom) for horizontal and vertical pump beam displacements, $\Delta X$ and $\Delta Y$, along the respective axis. For comparison, the dashed lines indicate the measured pump beam position.}
		\label{fig.4}
	\end{figure}
	We test for the response of the spatial photon intensity distribution in the double-well-system superimposed by the weak harmonic trap subject to different initial conditions, as realized by varying the pump configuration. These measurements present an important extension to our previous study of the thermalization of light in the same system carried out with a fixed, spatially centered pump beam \cite{Kurtscheid}. 
	
	Limitations of the dye-induced thermalization process in a harmonic potential provided by curved cavity mirrors have been investigated in previous work. In particular, the temporal equilibration dynamics of the photon gas have been intensively examined as a function of the coupling to the dye reservoir, revealing that at low dye absorption rates a thermal spectral distribution cannot be reached \cite{Julian1}. Further experiments investigated the condensate response to displacements of the pump region relative to the harmonic trap center, showing a spatial and energetic thermal redistribution despite the nonequilibrium initial state preparation up to radial displacements of up to \SI{60}{\micro m} from the trap center  \cite{JanThermal, JanCondensed,Nyman0}. 
	
	The present experiment examines the sensitivity of the fluorescence emission profile to pump beam displacements in our prototyped hybrid harmonic-double-well microcavity. At the longitudinal mode number $q=11$, the cutoff wavelength $\lambda_\mathrm{c}=\SI{586}{nm} $, and in Rhodamine 6G ethylene glycol dye solution (1\,mmol/l) with a refractive index of $n\approx 1.44$, the maximum imprinted deformation height of \SI{15}{nm} translates into a potential depth of \SI{14}{meV} corresponding to 0.55\,$k_\mathrm{B}T$ (parameters as in ref. \cite{Kurtscheid}). In order to spatially shift the excitation region in this potential, the pump laser beam of wavelength of \SI{532}{nm} and diameter of around \SI{70}{\micro m} is now scanned perpendicular and parallel to the double-well axis in two distinct measurement series respectively (Fig. 4b), while fixing the total photon number at $N=1.35N_\mathrm{c}$, where $N_\mathrm{c}$ denotes the critical photon number,  by correspondingly adjusting the pump beam power. Figures 4c and d show two key parameters for the spatial response of the photon gas fluorescence profile to the changes in the pump configuration. First, we find that for spatial displacements of the pump beam  $\Delta X$ and $\Delta Y$, orthogonal to and along the double well axis from the center, up to \SI{30}{\micro m}, the center of gravity of the emission nearly locks and decouples from the pump beam position (dashed lines in Fig. 4d) moving less by a factor of $\sim$35, as attributed to the thermalization in the harmonic trapping potential. Second, the analysis of the radial intensity profile shows that the full width at half maximum of the thermal cloud grows from $\Delta d_\mathrm{FWHM} \approx \SI{40}{\micro m}$ at central excitation by around \SI{15}{\micro m} (\SI{20}{\micro m}) for vertical (horizontal) displacements, indicating that photons pumped to highly excited modes with large off-center probability densities are not efficiently redistributed to near-axis modes in the center. We find this region of influence to be significantly reduced compared to previous work \cite{JanThermal},  as attributed to higher scattering losses in the present experiment where the mirror surface roughness of around \SI{1}{\angstrom}  exceeds the roughness of \SI{0.3}{\angstrom} of previous setups, as well as the here used different trapping potential. Nevertheless, we observe thermalization even for pump spots displaced significantly from the center of the trap.

	\section{Conclusions and outlook}
	We developed a novel technique to control the surfaces of Bragg mirrors via direct laser writing. With an intrinsic lateral resolution of \SI{3.6}{\micro m} point-like structures up to a height of around 50\,nm can be realized. Implementing an iterative structuring method enabled the creation of more complex two-dimensional mirror surfaces with mean deviations below \SI{2}{\angstrom} from the target profile masks. We observed no significant decrease in the mirror reflectivity up to deformation heights of 18\,nm. Within these limits, local elevation of the mirror surface allows us to set up a variety of trapping potentials for photon gases in the dye-filled microcavity setup. As an example, a double-well structure superimposed by a weak harmonic potential has been produced giving rise to a spatial superposition state as the energetic ground state. Thermalization of photons within such a potential landscape and condensation into the spatially split state has been observed \cite{Kurtscheid}. For the same potential structure, we have here studied the limits of the thermalization process based on the response of the spatial photon distribution to different pumping geometries. The presented deformation technique combined with the dye-cavity experimental setting enables the creation of novel variable potentials for two-dimensional photon gases.
	
	For the future, in the presence of effective photon-photon interactions the ground state in a suitable lattice geometry can become entangled and be populated in a thermal equilibrium process \cite{Kurtscheid}. Other prospects of the described method to generate variable potentials includes searches for superfluidity of the photon gas by studying the flow around imprinted perturbations, with potential implications for optical turbulence. Note that besides microscopic Kerr-nonlinear interactions, also temporally delayed thermo-optic interactions have been predicted to allow for superfluidity \cite{Hadiseh}. From a technical perspective, it would be interesting to investigate geometric modifications of the mirror layer design, to reach deformation heights on the order of the optical wavelength, which in addition to a wide class of microcavity applications would also e.g. open ways to holographic beam shaping.

	\begin{acknowledgments}
	We acknowledge funding by the Deutsche Forschungsgemeinschaft (DFG, German Research Foundation) under Germany's Excellence Strategy – Cluster of Excellence Matter and Light for Quantum Computing (ML4Q) EXC 2004/1 – 390534769, the DFG within the focused research center SFB/TR 185 (277625399), the EU within the ERC Advanced Grant project INPEC and the Quantum Flagship project PhoQuS, and the DLR with funds provided by the BMWi (50WM1859). J.S. acknowledges support from Churchill College, Cambridge.
	\end{acknowledgments}

\bibliographystyle{apsrev4-1}
\bibliography{bib}

\end{document}